\documentstyle[prl,aps]{revtex}

\begin{document}

\newcommand{\bib}{\bibitem}
\newcommand{\er}{\end{eqnarray}}
\newcommand{\br}{\begin{eqnarray}}
\newcommand{\be}{\begin{equation}}
\newcommand{\ee}{\end{equation}}
\newcommand{\epe}{\end{equation}}
\newcommand{\bea}{\begin{eqnarray}}
\newcommand{\eea}{\end{eqnarray}}
\newcommand{\ba}{\begin{eqnarray}}
\newcommand{\ea}{\end{eqnarray}}
\newcommand{\epa}{\end{eqnarray}}
\newcommand{\ar}{\rightarrow}

\def\r{\rho}
\def\D{\Delta}
\def\R{I\!\!R}
\def\l{\lambda}
\def\D{\Delta}
\def\d{\delta}
\def\T{\tilde{T}}
\def\k{\kappa}
\def\t{\tau}
\def\f{\phi}
\def\p{\psi}
\def\z{\zeta}
\def\hx{\widehat{\xi}}
\def\na{\nabla}

\begin{center}

{\bf General Relativity as a (constrained) Yang-Mills's Theory and
a Novel Gravity with Torsion.}

\vspace{1.3cm} M. Botta Cantcheff\footnote{e-mail: botta@cbpf.br}

\vspace{3mm} Centro Brasileiro de Pesquisas Fisicas (CBPF)

Departamento de Teoria de Campos e Particulas (DCP)

Rua Dr. Xavier Sigaud, 150 - Urca

22290-180 - Rio de Janeiro - RJ - Brazil.

\end{center}

\begin{abstract}
We show that General Relativity (GR) with cosmological constant
{\it may be formulated} as a rather simple
 constrained $SO(D-1, 2)$ (or $SO(D, 1)$)-Yang-Mills (YM) theory. Furthermore, the spin connections of the
  Cartan-Einstein
 formulation
  for GR appear as solutions of a genuine $SO(D-1,1)$-YM.

 We also present a theory of gravity with torsion
   as the most natural extension
    of this result. The theory comes out
     to be strictly an YM-theory upon relaxation of a
     suitable constraint. This work sets out to enforce the close connection
   between YM
   theories and GR by means of a new construction.

\end{abstract}
\section{Introduction}

      There exists a great deal of attempts to formulate GR as a YM-type theory or, in general,
as a gauge theory
\cite{utiy,kibble,sciama1,sciama2,rmp71,mm,sw0,obr,refobr,gotz,hehl,gor}.
However, there is
 not yet a
simple and conclusive result that establishes this connection very
neatly.

Hehl et al \cite{hehl} consider the Poincare group as the local
symmetry group, and the basic dynamical variables of GR are
obtained from the gauge fields (the connection) on a principal
bundle over spacetime.

Mac Dowell and Mansouri \cite{mm} proposed a gauge theory of
gravity based on the group $SO(3,2)$ (or $SO(4,1)$), for the first
time. The Poincare group is obtained by the Wigner-In\"on\"u
contraction \cite{gil}.

 Other authors, Stelle-West \cite{sw0} and Gotzes-Hirshfield \cite{gotz}
  worked also along the same stream, in particular Stelle-West \cite{sw0},
    recover GR by imposing a constraint in the action explicitly.

There has been an
       increasing revival and interest in this type of formulation in order to find
        11d-SUGRA from algebras
        corresponding to
        12-dimensional theories (F-theories) \cite{sw,van,sugrad4,sugra12}.

These approaches, however, have some disadvantages:

{\bf .} They are gauge formulations, but they do not have a {\it
genuine} YM-structure.

 This means that the equations
 are of the Maxwell type, i.e they derive from an action proportional
 to the square of the gauge field-strength two-form. This is an important fact in order to implement a
 universal quantization scheme,
 similar to the one adopted for the other forces of Nature.

{\bf .} They are restricted to D=4.

{\bf .} They do not show how a genuine {\it YM-current} must be
consistently related to the energy-momentum tensor.

In our work, we focus our attention on
      the equations of motion (of a YM's theory) rather than insisting on an analysis based upon the action.
We start off with equations manifestly different from the ones of
the approaches referred to above.

 This approach succeeds in solving the points mentioned above. Furthermore, it shows clearly {\it where is}
 the real difference between GR and an YM-theory; this relies on a single
 constraint which has an extremely simple interpretation: the torsion-free condition.

Remarkably, this constraint reduces a second order theory (YM) to
a first order one, the so-called Einstein-Cartan
 formulation of GR. The YM structure assumes an
  internal symmetry group which is broken by that constraint, leading 
to
$SO(1,3)$ as the internal residual gauge group.

This paper is organized according to the following outline:

     In Section 2, we briefly introduce the Einstein-Cartan formalism and the main
      definitions of an YM-structure are established; next, in Section 3, it is shown that
spin connections are solutions of $SO(D)$-YM, with
 an interesting form for the sources.
Calculations in this direction have been done, in a different
context, mainly for the purpose of numerical calculations
 \cite{pir,klain,vp}.

The main result, the YM-formulation of GR, is presented in Section
4, where an appealing formulation of a theory of gravity with
torsion arises naturally.

Finally, our Concluding Remarks are collected in Section 5.

 The results presented below hold for an arbitrary $D$, but we particularize for $D=4$ and a Lorentzian signature to
  have in mind Einstein-Cartan GR-theory, though it is not necessary.

\section{Einstein-Cartan formalism for GR and the spin connection as an YM variable.}

In this work, we shall use the abstract index notation
\cite{wald}; namely, a tensor of type $(n,m)$ shall be
 denoted by $T^{a_1 .....a_n}_{b_1 .....b_m}$, where the latin index stand for the numbers and types of variables
  the tensor acts on and not as the components emelves on a certain basis. Then, this is an object having a
   basis-independent meaning.
 In contrast, greek letters label the components, for example $T^{\mu \nu}_{\alpha}$
 denotes a basis component of
  the tensor $T^{ab}_c$. We start off with the Cartan's formalism of GR.
 We introduce \cite{wald} an orthonormal basis of smooth vector fields $(e_{\mu})^a$,
 satisfying
\be (e_{\mu})^a(e_{\nu})_a = \eta_{\mu \nu},\label{orto}
 \ee where
$\eta_{\mu \nu}=diag(-1,1,1,1......,1)$. In general, $(e_{\mu})^a$
is referred to as {\it vielbein}.
 The metric tensor is expressed as
\be\label{defg-e} g_{ab}=(e^{\mu})_a(e^{\nu})_b \eta_{\mu \nu}. \ee
From now, component indices $\mu,\nu,..$ will be raised and lowered 
using the flat metric $\eta_{\mu\nu}$ and the the abstract ones, $a,b,c...$
with space-time metric $g_{ab}$.

 We define
now the {\it Ricci rotation coefficients}, or {\it
spin-connection}, \be (w_{\mu\nu})_a=(e_{\mu})^b\nabla_{a}(e_{\nu})_b
, \label{defw} \ee where $w_{a\mu\nu}$ is antisymmetric which,
together with (\ref{orto}, is equivalent to the compatibility
condition
 \be\nabla_a g_{bc}=0\label{compat}\ee. From (\ref{defw}), we have
\be \nabla^{}_{a}~e^{\mu }_{~~b}=-w^{\mu\nu}_{~~~a} e^{}_{\nu \, b}. \label{e-w}
\ee

 Taking the antisymmetric part, \be \nabla^{}_{[a}~e_{~~b]}^{\mu
}=-w^{\mu\nu}_{~~~[a} e^{\alpha}_{~~b]}\eta_{\nu \alpha}. \label{antis-e-w} \ee
We have adopted the convention of anti-
symmetrization:$(...)_{[ab]} = ((...)_{ab} - (...)_{ba})/2$.

In the original Einstein's formulation of GR, the

connection is assumed to be torsion-free, this is expressed by:
\be
\partial^{}_{[a}~e_{~~b]}^{\mu
}=-w^{\mu\nu}_{~~[a} e^{\alpha}_{~~ b]}\eta_{\nu \alpha}. \label{notorsion}
\ee

 The components of the Riemman's tensor in this orthonormal
basis are given as follows \be R^{~~\mu \nu}_{ab}:= 2\partial^{}_{[a}
w^{\mu \nu}_{~~b]} + 2w^{\mu \rho}_{~~[a}w^{\sigma\nu}_{~~b]}
\eta_{\rho\sigma}. \label{p1} \ee Equations (\ref{notorsion}) and
(\ref{p1}) are the {\it structure equations} of GR  in Cartan's
framework.

The Einstein's equation is \be e_{\mu}^{~~a} R^{~~\mu \nu}_{ab} =
\kappa ~ e_{\nu}^{~~a} T'_{ab}e,\label{eins} \ee where one has defined $T'_{ab} :=
T_{ab} + g_{ab} (T_{cd}g^{cd})/2$, $T_{ab}$ being the energy
momentum tensor.

Equations (\ref{e-w}) and (\ref{eins}) sets up a system
 of coupled first-order non-linear equations for the variables $(e,w)$
which determine \footnote{Together with 
the antisymmetry condition for $w_a$.} the dynamics of GR. Metric and covariant derivative result finally
 defined in terms
of these variables as seen from (\ref{defg-e}) and (\ref{e-w}).

 This yields the so-called "Einstein-Cartan formalism"; we obtain thereby a first order
 Einstein-Hilbert's action which
  can be expressed as
\be S={1 \over 2\kappa^2}\int dx^D e R^{~~\mu \nu}_{ab}e_{\mu}^{~~a} e_{\nu}^{~~b}, \label{E-H}
\ee where $e=(-det~ g)^{1/2}= det ~(e^{\mu}_{~~a})$. If we wish consider
a non-vanishing cosmological constant, $\Lambda$,
 $R^{~~\mu \nu}_{ab}$ must be replaced by
\be R^{~~\mu \nu}_{ab} + \Lambda e^{[\mu}_{~~a} e^{\nu]}_{~~b}.
\label{cosmol} \ee
Finally, recall the YM's equations for a generic $SO(p,q)$ gauge field
 $A^{AB}_a$, where $A,B=1,....p+q$
label on the components of the gauge field and $A^{AB}_a$ is a $(p+q)(p+q-1)/2$
 collection of one forms ($A^{AB}_{~~a}=-A^{BA}_{~~a}$); They are the the dynamical
variables of the theory whose equations of motion are second order. 

We low and rise these internal index with the flat metric $\eta_{AB}$, which has
 $p$ and $q$ eigenvalues being $1$ and $-1$ respectively.

 Let us
define the field-strength: \be F^{~~AB}_{ab}:= 2\partial^{}_{[a}
A^{AB}_{~~b]} + 2A^{AC}_{~~~[a}A^{DB}_{~~~b]} \eta_{CD}, \label{fs} \ee

 In a general curved Einstein's spacetime (with canonical connection $\nabla_a$),
 the YM's equations are second order in the potentials:
\be \nabla^a F^{~~AB}_{ab} + 2 A^{C[A\,a} F^{~~B]}_{ab~~C} = J^{~~AB}_b
\label{yme} \ee
 where $J^{~~AB}_b$  is the YM current.
 It is straightforward to show that this equation derives from a typical YM action, proportional to $F^2$.

This equation can be written shortly as \be {\cal D}^a F^{~~AB}_{ab}
= J^{~~AB}_b, \label{div} \ee where we have defined  the $SO(D)$
covariant derivative \be {\cal D}_a K^{A_1....A_n}= \nabla_a
K^{A_1...A_n} + \Sigma_{i=1}^n A^{A_i}_{~~C\,a}~ K^{A_1...A_{i-1} C
A_{i+1}...A_n}, \label{covdef} \ee

$K^{A_1...A_n}$ being a spacetime tensor of arbitrary rank\footnote{ Notice that
 two covariant derivatives appear: the background-covariant
  derivative, $\nabla_a$ and the YM's one,
  ${\cal D}_a$. Actually, ${\cal D}_a$ can be thought as the single one;
   recalling that $\nabla_a$ 
    acts on spacetime (abstract) indices, the Christoffel symbols shall be
 taken into account.}.

Remark: Equation (\ref{fs}) together with (\ref{yme}) constitute
the full structure of an $SO(p,q)$-YM theory up to a gauge-fixing,
 i.e; since the gauge invariances of equations (\ref{div}),
 the fields $A$ are not fully determined by these ones; thus,
in order to solve a YM equation system, an additional (gauge)
  condition needs to be imposed.

We demonstrate below an important claim for this paper: the spin
connection sector of the GR-solutions are solutions
 of a current YM theory too.
This result will be critical for the construction of the next section.

\vspace{7mm}
Proposition {\bf(2.1)}: The spin connection $w_a$ of a
$D$-dimensional smooth oriented Einstein-Cartan space time \footnote{Of signature $(-,+,+,+)$.},
constitutes an
 $SO(D-1,1)$-gauge field satisfying the $SO(D-1,1)$-Yang Mills equations
 (\ref{fs}),(\ref{yme}) on this (curved) space
  time. 
\vspace{7mm}
In other words, this means that if $w^{\mu\nu}_{~~~a}$ is an antisymmetric field
 defined in terms of 
vielbein fields and covariant (compatible) derivative by (\ref{defw}) such that the 
Einstein Equation (\ref{eins}) are satisfied, thus, the Yang-Mills equations
 (\ref{fs}),(\ref{yme})
 on the corresponding space time, hold
for the gauge field ($A$) taken to be $w^{\mu\nu}_{~~a}$.

Proof:

We have the field strength for the spin connection field defined
as in (\ref{fs}): \be R^{~~\mu \nu}_{ab}:= 2\partial^{}_{[a} w^{\mu
\nu}_{~~b]} + 2w^{\mu \rho}_{~~[a}w^{\sigma\nu}_{~~~b]}
\eta_{\rho\sigma}, \ee
  which is again an $SO(D-1,1)$ gauge invariant object, but now $D$ agrees with the dimension
 of the spacetime (supposed
   to be a Lorentzian one). Henceforth, let us fix $D=4$.

 The Einstein-Cartan's equations will describe a subset of solutions of a YM theory, which has not invariance.

The Bianchi identity reads \be \nabla^a R_{bcda} +
\nabla_{[b}R_{c]d}=0, \label{bianchi1} \ee
 but, using the symmetry properties of the Riemman's tensor $R_{bcda}=R_{dabc}=-R_{cbda}$, we find
\be \nabla^a R_{adbc} - \nabla_{[b}R_{c]d}=0 \label{bianchi2} \ee
 Einstein's equation can be writen as $R_{ab}=\kappa T'_{ab}$, where  $T'_{ab} := T_{ab} + g_{ab} (T_{cd}g^{cd})/2$ has
  been defined. Finally, we found an equation which holds for the on-shell GR \cite{vp}:
\be \nabla^a R_{adbc} - \kappa\nabla_{[b}T'_{c]d}=0. \label{bianchiT}
\ee On the other hand, the Riemann tensor is related to the
YM-type field strength by \be R_{ad}^{~~~\mu \nu} =R_{adcb} e^{\mu\,c}
e^{\nu \,b}; \label{p2} \ee taking the divergence, it yields: \be
\nabla^a R^{~~~\mu \nu}_{ad}:= [\nabla^a \,R_{adcb}]e^{\mu \,c}
e^{\nu \,b} + R_{adcb} [\nabla^a e^{\mu \,c} e^{\nu \, b}]\label{p3}
\ee

Replacing (\ref{bianchiT}) at the R.H.S. of this equation, \be
\nabla^a R^{~~~\mu \nu}_{ad}=  \kappa [\nabla^{}_{[c}T'_{b]d}] e^{\mu \, c}
e^{\nu \,b} + R_{adcb} [\nabla^a e^{\mu\, c} e^{\nu \,b}].\label{p4}
\ee 

Let us concentrate on the last term; using the antisymmetry in $c,b$
for the Riemann tensor, we may write: \be R_{adcb} [\nabla^a
e_{\mu}^{~~c} e_{\nu}^{~~b}] =R_{ad  \alpha \beta} e^{\alpha}_{~~c}
e^{\beta}_{~~b}[\nabla^a  e_{\mu}^{~~c} e_{\nu}^{~~b}]; \label{p5} \ee
  but, with the help of (\ref{defw}),
\be
 e_{~~c}^{\alpha} e_{~~b}^{\beta}\nabla^a e^{~~c}_{\mu} e^{~~b}_{\nu} =  w^{\alpha~~a}_{~\mu}~\delta^{\beta}_{~\nu}
  + w^{\beta~~a}_{~\nu}~ \delta^{\alpha}_{~\mu}.\label{p6}
\ee Replacing this in (\ref{p5}), \be R_{adcb} [\nabla^a e^c_{\mu}
e^{~~b}_{\nu}] = w^{\alpha~~a}_{~\mu} ~R_{ad \alpha\nu} + w^{
\alpha~~a}_{~\nu}~R_{ad \alpha\mu} .\label{p9} \ee Finally,
substituting it in (\ref{p4}), we have the remarkable result:

\be \nabla^a R_{ab\mu \nu} -  w^{\alpha~~a}_{~\mu}~R_{ad
\alpha\nu} - w^{\alpha~~a}_{~\nu}~ R_{ad \alpha\mu} = j^{~~\mu
\nu}_{b},  \label{p10} \ee which has the form of the typical YM
equation (\ref{yme}) with the "YM-current" defined as \cite{vp}:
\be j^{~~\mu \nu}_{a} :=  \kappa [\nabla_{[c}T'_{b]a}] e^{~~c}_{\mu} e^{~~b}_{\nu}
\label{p11} \ee This completes the demonstration.

\section{Equivalence of GR to a (constrained) $SO(3;2)$ (or $SO(4;1)$)-YM
 Theory and inclusion of torsion.}

 The aim of this section is to show that GR (with cosmological constant)
  {\it can be written as} a YM theory plus certain constraints whose elimination led to
   a natural way to define the theory (GR) including torsion. This is the main construction
   of this paper.

    We shall restrict
   ourselves to empty space to render more clear and evident some points; but the generalization
    to the case when
    matter is taken into account is straightforward and will be done at the end of the section.

 Let us define this theory and prove that this is equivalent to GR.

 Let $M$ be a four-dimensional manifold with a smooth (oriented) metric
$(M,g_{ab})$ of signature $(-,+,+,+)$. We shall also asume that,
to each point $p \in M$, we can assign a real 5-dimensional vector space, $V$,
equipped with a scalar product
 given by $\eta_{CD} :=diag(-1,1,1,1,(-1)^s)$.
This defines the group $SO(3+s;2-s)$ (where $s=\pm 1$ ), since this 
is the group that preserves the structure on $V$.

  The dynamics is fully described by second order equations
\footnote{The same ones that (\ref{yme}) and (\ref{fs}) with $d=4$, $A;B=1,....5$.}
in the gauge variables $A^{AB}_a$ (which for definition, are antisymmetric in $A,B$):
\be \nabla^a G^{~~AB}_{ab} + 2 A^{C[A\,|a} G^{~~B]}_{ab~\,C} = J^{~~AB}_b,
\label{yme5} \ee where the field strength reads: \be G^{~~AB}_{ab}:=
2\partial_{[a} A^{AB}_{~~b]} + 2A^{AC}_{~~~[a}A^{DB}_{~~~b]} \eta_{CD}.
\label{fs5} \ee

Note that for writing down these equations we have implicitly supposed
the existence of a Riemannian metric $g$
and a covariant derivative $\nabla$. Without them, equations
(\ref{yme5}) would not make sense, however, below we will close the system
by imposing relations such that
this geometry structure will be given in terms precisely of the gauge variables. 

    Alternatively, it is more convenient to express this structure in the language described
     at the previous section,
    we are
    assuming the existence of $(e,w)$, where $e$ is defined trough
    \be g_{ab}=(e^{\mu})_a(e^{\nu})_b \eta_{\mu \nu}, \ee

     and the spin-connection coefficients, $(w_{\mu\nu})_a$, are defined in the general case,
 i.e for
     {\it any} covariant derivative:

    \be (w^{}_{\mu\nu})_a=(e_{\mu})^b\nabla^{}_{a}(e_{\nu})_b
, \label{defw2} \ee.

     Then, we can define the torsion\footnote{Notice that additional
 structure as the antisymmetry of the one forms,
 $w_a$, which implies the compatibility
  condition (\ref{compat}), together with
the current torsion-free condition are not introduced {\it a priori} in this formulation,
 they shall be get from the equations defining the theory.}
 by
\be
\theta^{~~\mu}_{ab}:= \partial_{[a}e_{~~b]}^{\mu } + w^{\mu\nu}_{~~[a}
e_{~~b]}^{\rho}\eta_{\nu \rho}\label{deftor} .\ee

 Now, we make a {\it global} choice of the fifth basis element of $V$: $U$, defined satisfying,
  $\nabla_a U=0$. Then, we
  define:

\be
 E_{~~b}^A := A_{~~\,b}^{A5}=A^{A}_{~B~b}~ U^B . \label{defE}
\ee
  Let us use the greek letters to denote the first four components, ie $A=\mu,5$, with $\mu=1,...4$.

The introduction of that vector is related to the Wigner-In\"on\"u
 contraction which reduces $SO(3+s;2-s)$ to $ISO(3,1)$, the
  standard gauge group of GR \cite{mm,sw0,hehl}. Notice, however,
 that a contraction parameter has not
 yet been introduced
  and this will be not necessary in this construction.

Then, as previously announced, we write down the suplementar
condition (constraint)
 \be
  G^{~~A\,5}_{ab}=0. \label{Anotorsion}
   \ee

This is a first order equation relating the gauge fields; then, it
constitutes a constraint for the above YM dynamical system. Notice
that up this point, the theory, namely the YM-equations plus
(\ref{Anotorsion}), manifestly appears to be $SO(1,3)$-gauge
invariant.

Replacing (\ref{Anotorsion}) in the $A-5$ component of (\ref{yme5}) -with
$J^{~~AB}_b=0$-, \be A^{C[A\,|a} G^{~~B]D}_{ab} \eta_{CD}=0; \ee taking
$A=\mu$ and $B=5$, we find \be [A^{ C \mu \, a} G^{~~5 D}_{ab}-A^{C 5\,a}
G^{~~\mu D}_{ab}] \eta_{CD}=0. \ee Since $\eta_{CD}$ is diagonal,
using again (\ref{Anotorsion}), we obtain \be G^{~~\mu
\nu}_{ab}A_{\mu 5}^{~~a} =0. \label{Apreg} \ee

Finally, wqe shall relate the geometry variables with the YM-fields.
Actually, this YM-type theory, coincides with GR once the
identifications \cite{mm,sw0,hehl}
 are imposed:
\be \mu e^{\mu}_{~~a} = E^{\mu}_{~~a}, \label{ide} \ee

 \be
  w^{\mu
\nu}_{~~~a}=A^{\mu \nu}_{~~~a}, \label{idw}
 \ee
 where $\mu$ is a parameter which has inverse length dimension
 related to the cosmological constant,
as it shall become clearer later on. This parameter is introduced in order to give a dimensionless
vielbein field, however, there is no some indication a priori of any scale of length in the theory.

From (\ref{idw}), $w$ must be antisymmetric.
 This identification finally fixes
the relation between the fields of theory ($A^{AB}_{a}$) and the background
space time structure.  Equation (\ref{Anotorsion}) is recognized
as the first of the Cartan's structure equations,
  which expresses the non-torsion condition.

Replacing (\ref{ide}) into (\ref{Anotorsion}), $A$ (or $w$) can be
solved in terms of $e_b^{\nu}$, in an Einstein-Cartan scheme.
Then, in according with (\ref{deftor}), we deduce that the torsion
of the spacetime covariant derivative $\nabla$, vanishes. This,
plus the antisymmetry of $w$ determines completely this
connection, which results to be the canonical one.

 Equation
  (\ref{Apreg})
 with the identifications(\ref{idw}), (\ref{ide}), read as the (vacuum) Einstein's equation:
\be G^{~~\mu \nu}_{ab}e_{\mu}^{~~a} =0 \label{preg} \ee This is the
(vacuum) Einstein's equation with a cosmological-constant term
because \be G^{~~\mu \nu}_{ab}:= 2\partial_{[a} w^{\mu \nu}_{~~b]} +
2w^{\mu \rho}_{~~[a}w^{\sigma\nu}_{~~b]} \eta_{\rho\sigma} +
 2(-1)^s \mu^2 e^{[\mu}_{~~a} e^{\nu]}_{~~b}\label{GR1}
\ee ie \be G^{~~\mu \nu}_{ab}:= R^{~~\mu \nu}_{ab} + 2(-1)^s \mu^2
e^{[\mu}_{~~a} e^{\nu]}_{~~b}, \label{GR2} \ee this resembles
(\ref{cosmol}) with $\Lambda = (-1)^s \mu^2 $ .

Notice furthermore that this (the Einstein's theory), is all the
structure we can extract of the theory, in other words, the other
YM equations do not introduce extra conditions. To show this, we
shall use strongly the result of the previous section.

Going to the $\mu,\nu$-components of (\ref{yme5}), we get \be
{\cal D}^a R^{~~\mu \nu}_{ab} + (-1)^s m^2 {\cal D}^a (e^{[\mu}_{~~a}
e^{\nu]}_{~~b})=0  \label{rg1} \ee

Notice that by virtue of (\ref{idw}) the full covariant divergence
in (\ref{yme5})
 agrees with the one of the proposition {\bf(2.1)} ($SO(1,3)$)
\be {\cal D}^a G^{~~\mu \nu}_{ab} = \nabla^a G^{~~\mu \nu}_{ab} + 2
A^{C[\mu\,|a} G^{~~\nu]}_{ab ~C} \ee using (\ref{Anotorsion}), \be {\cal
D}^a G^{~~\mu \nu}_{ab} = \nabla^a G^{~~\mu \nu}_{ab} + 2
A^{\alpha[\mu\,|a} G^{~~\nu]}_{ab~\,\alpha} . \ee

The term ${\cal D}^a  (e^{[\mu}_{~~a} e^{\nu]}_{~~b})$ vanishes using
(\ref{defw2}) -or equivalently, (\ref{e-w})-.

Thus, (\ref{GR2}) reduces to the $SO(3,1)$-YM equation, which
already has been proven
 -proposition {\bf(2.1)}- to be
 identically satisfied by the fields $e,w$ being solutions of the Einstein equation (empty)
  in the presence of a
 cosmological constant):
\be R^{~~\mu \nu}_{ab}e_{\mu}^{~~a} = \kappa T'_{(\Lambda)
ab}e^{\nu\,a},\label{eins2} \ee which completes the proof of our
claim.

The generalization to a non-trivial energy-momentum tensor,
$T_{ab}$, is straightforwardly
 obtained by starting with an
YM-theory with sources. In order to recover GR, the YM-current
must be defined in terms of the general
  energy-momentum tensor:
\be J^{~~\mu }_{a 5} := \kappa T_{a}^{'~~\mu}, \label{fuentegen} \ee where
$T_{a}^{'~\mu}=T'_{ab}e^{\mu \, b}$.

Finally, we have consistency with the above results if the other
components of the YM-current
 are defined to satisfy:
\be J^{~~\mu \nu}_{a} :=  \kappa [\nabla_{[c}T'_{b]a}] e^{~~c}_{\mu} e^{~~b}_{\nu}.
\label{fuente2} \ee

\subsection{ A gravity theory with torsion.}

Torsion appears in a natural way in modern formulations of the
gravitational theories \cite{shap}. This supports the framework
discussed below (our final result).

Notice that, by relaxing the constraint
(\ref{Anotorsion}), we are naturally led to a particularly elegant theory of
gravity
 {\it with torsion}, which remarkably enough turns out to be an ordinary $SO(3+s;2-s)$-YM.
This theory is described as before by the dynamical equations:

\be \nabla^a G^{~~AB}_{ab} + 2 A^{C[A\,|a} G^{~~B]}_{ab~\,C} = J^{~~AB}_b,
\label{yme5t} \ee where
 \be G^{~~AB}_{ab}:= 2\partial_{[a}
A^{~~AB}_{b]} + 2A^{AC}_{~~~[a}A^{DB}_{~~~b]}~ \eta_{CD}, \label{fs5t} \ee

\be J^{~~AB}_b=(J^{~~\mu 5}_b ; J^{~~\mu
\nu}_b)=\kappa (T_{a}^{'~~\mu}; \,[\nabla_{[c}T'_{b]a}] e^{~~c}_{\mu}
e^{~~b}_{\nu}).\ee

  In order to describe
gravitation, the identification constraints to be imposed are
(\ref{ide}), (\ref{idw}); thus, the physical spacetime torsion is
given by \be \Theta:= \mu^{-1} G^{~~\mu 5}_{ab}, \label{deftor2}\ee

where we observe that the cosmological constant must be
non-vanishing.

 The modified
Einstein's equation results from the $\mu-5$ component of
(\ref{yme5}). It reads:
 \be
R^{~~\mu \nu}_{ab}e_{\mu}^{~~a} = - {\cal D}^a \Theta^{~~\mu}_{ab} +
\kappa~T^{'~~\mu}_{b}, \label{mu5}
  \ee

  The components $\mu-\nu$ of Equation constitute complementary ones, which are
 identities when
  the torsion is vanish.

\section{Concluding remarks.}
We conclude by stressing a remark: the meaning of the
identification expressed by equations (\ref{ide}), (\ref{idw}).
Formally, such an identification shall be looked upon as a constraint.

It has been argued in similar approaches that one 
 can formulate GR without cosmological constant, by setting
 the appropriate
  limit $\mu \to 0$;
  in this case, the YM-group tends  remarkably to the Poincare-Lorentz
 one via the well-known algebra
   contraction. Care is needed with this since point: in this limit,
the structure underlying this approch appears to be singular as we can see in
 equations (\ref{ide})
and (\ref{deftor2}). These are two important issues and commonly,
they are not remarked in the previous similar formulations.

   It remains to be more deeply investigated the existence of exact solutions to YM
 theories starting from the particular ones well-known in GR. The issue of quantizing
the theory in the presence of the constraint in the form presented here, is also a
 delicate and relevant matter to be
pursued.

{\bf Aknowledgements}: The author is indebted to Francesco Toppan
for suggesting this interesting research line.  J. A. Helayel-Neto
is aknowledged for many invaluable discussions and pertinent
corrections on an earlier manuscript. Thanks are due also to Dr.
Andrew Waldron for helpful and
 relevant comments and criticism. The GFT-UCP is aknowledged for the kind hospitality.
CNPq is acknowledged for the invaluable financial help.

\end{document}